\documentclass[aps,prl,twocolumn,preprintnumbers,amsmath,amssymb,superscriptaddress,10pt]{revtex4-2}
\usepackage{amsmath,graphicx}
\usepackage[utf8]{inputenc}
\usepackage[T1]{fontenc}
\usepackage{xcolor}
\usepackage{textcomp}
\usepackage{bm}
\usepackage{physics}
\usepackage{amsmath}
\usepackage{tikz}
\usepackage{mathdots}
\usepackage{yhmath}
\usepackage{cancel}
\usepackage{color}
\usepackage{array}
\usepackage{multirow}
\usepackage{amssymb}
\usepackage{gensymb}
\usepackage{tabularx}
\usepackage{extarrows}
\usepackage{booktabs}
\usetikzlibrary{fadings}
\usetikzlibrary{patterns}
\usetikzlibrary{shadows.blur}
\usetikzlibrary{shapes}

\usepackage{color}
\definecolor{LinkColor}{rgb}{0.75,0.0,0.2}

\usepackage{hyperref}
\hypersetup{
	pdfauthor={good guys},
	pdftitle={good title},
	colorlinks=true,
	citecolor=LinkColor,
	linkcolor=LinkColor,
	urlcolor=LinkColor,
}

\usepackage{listings}
\definecolor{lightgray}{gray}{1}

\usepackage{theorem}

\usepackage{tikz}

\newcommand{\nc}{\newcommand}
\nc{\braoprket}[3]{\langle#1|#2|#3\rangle}
\nc{\opn}[1]{\operatorname{#1}}
\nc{\avg}[1]{\langle#1\rangle}
\nc{\ketbrasame}[1]{|#1\rangle\!\langle#1|}
\nc{\swap}{\opn{SWAP}}
\nc{\E}{\mathbb{E}}
\nc{\Var}{\opn{Var}}
\nc{\dg}{\dagger}

\usepackage[normalem]{ulem}

\begin{document}

\title{Critical Topology: A New Perspective for Phase Transitions}

\author{Xue-Jia Yu}
\email{xuejiayu@eitech.edu.cn}
\affiliation{Eastern Institute of Technology, Ningbo 315200, China}

\begin{abstract}


\end{abstract}

\maketitle

Phase transitions and critical phenomena are among the most fundamental phenomena in nature, spanning a broad range of fields in physics, from cosmology and particle physics to condensed-matter physics and quantum information science. In traditional statistical physics textbooks~\cite {landau2013statistical,Sachdev_1999,cardy1996scaling}, Landau–Ginzburg theory provides the conventional framework for describing continuous phase transitions accompanied by spontaneous symmetry breaking, in which the order parameter evolves \emph{continuously} from zero to a finite value. At a critical point, the correlation length diverges, rendering microscopic lattice details irrelevant at long distances and allowing the critical point to be described by a universal low-energy field theory. Building on this framework, the renormalization-group (RG) theory developed by Wilson~\cite{Wilson1983RMP} demonstrated that different microscopic lattice models can flow toward the same fixed point in the low-energy limit, thereby explaining the origin of universality in critical phenomena. Together, the Landau–Ginzburg and Wilson RG theories form the standard paradigm for understanding phase transitions and critical phenomena, commonly known as the Landau–Ginzburg–Wilson (LGW) paradigm. A central goal of modern statistical physics is to classify phase transitions and identify new universality classes. Within the LGW framework, each universality class is characterized by a set of critical exponents, a principle that has guided theoretical, numerical, and experimental investigations of critical phenomena for decades. 

However, not all continuous phase transitions can be described within the LGW framework. Such transitions are commonly referred to as non-Landau transitions, with topological phase transitions being a prominent example~\cite{Kosterlitz2017RMP,Haldane2017RMP}. In the existing literature, the term topological phase transition is used in two main senses. The first refers to transitions driven by topological defects, such as the Kosterlitz–Thouless (KT) transition associated with the binding and unbinding of vortices. The second refers to continuous transitions between topological trivial to nontrivial phases~\cite{Wegner1971,Kogut1979RMP,Hasan2010RMP,Qi2011RMP}. Unlike conventional Landau–Ginzburg continuous transitions, topological phase transitions do not involve spontaneous symmetry breaking; instead, they are characterized by \emph{discrete} topological invariants, such as winding numbers and Chern numbers. These invariants are generally well-defined only when the bulk is gapped and become ill-defined when the bulk energy gap closes at a quantum critical points (QCPs). This has fostered a long-standing belief in condensed-matter physics: once the bulk energy gap closes, topological edge states delocalize and merge into the bulk, thereby losing their stability. In this sense, the stability of topological edge states relies on the existence of a bulk energy gap. Accordingly, both short-range-entangled topological insulators~\cite{Hasan2010RMP,Qi2011RMP} and symmetry-protected topological (SPT) phases~\cite{Wen2017RMP,Senthil2015ARCMP}, as well as long-range-entangled intrinsic topological orders~\cite{Wen2013ISRN,Wen2016NSR,Sachdev_2019}, have traditionally been studied within the framework of gapped systems.

These considerations reveal an apparent contradiction between criticality and topology: continuous quantum critical points are gapless, whereas topology has traditionally been formulated in terms of gapped quantum systems; criticality is characterized by continuously varying order parameters, whereas topology is described by discrete topological invariants. Consequently, for decades, criticality and topology—two central themes of statistical and condensed matter physics—have largely been regarded as fundamentally incompatible, developing along separate research trajectories with little overlap.

\begin{figure}[t]
    \centering
   \includegraphics[width=\linewidth]{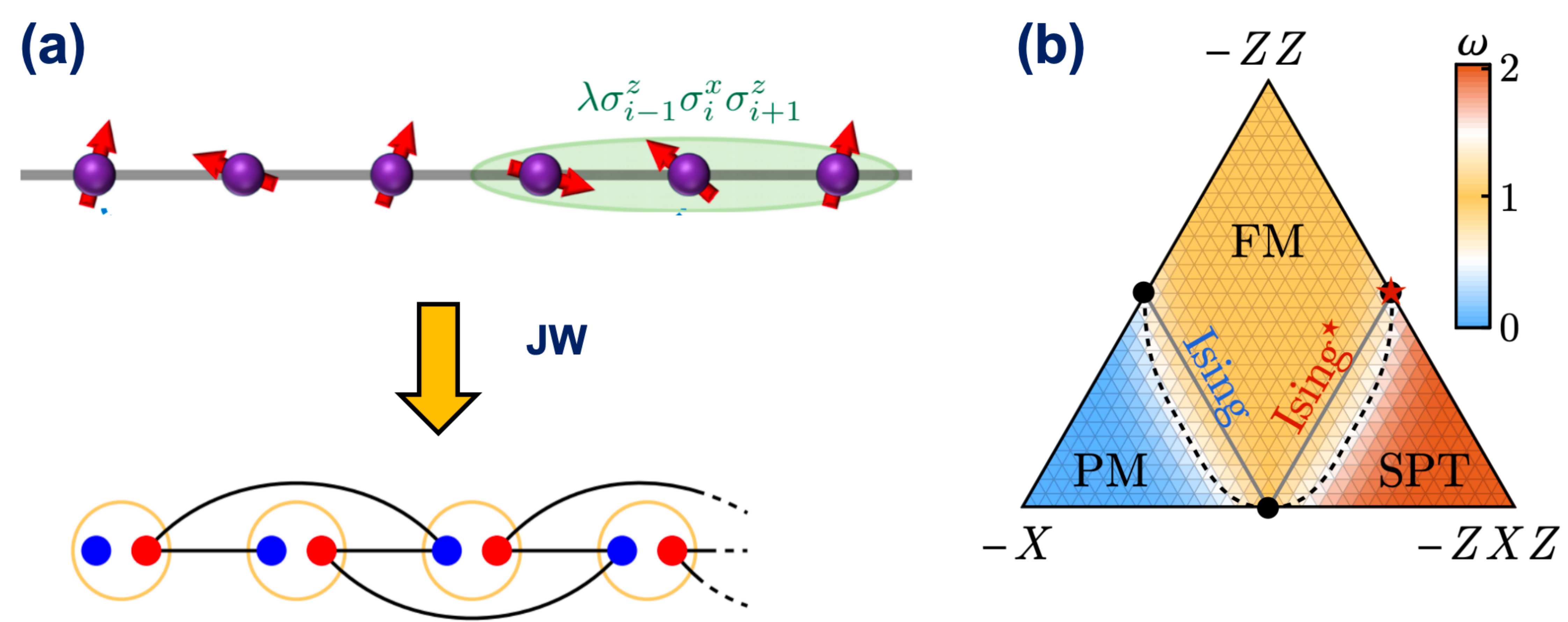}
    \caption{(a) Mapping of a one-dimensional spin chain to a Majorana-fermion model via the Jordan–Wigner (JW) transformation~\cite{Verresen2018PRL}. Orange circles denote lattice sites, while blue and red dots represent the two types of Majorana fermions. (b) Ground-state phase diagram of the one-dimensional extended cluster–Ising model~\cite{Duque2021PRB}. FM, PM, and SPT denote the ferromagnetic, paramagnetic, and symmetry-protected topological phases, respectively. Ising and Ising* label the ordinary and symmetry-enriched Ising critical lines, respectively.}
    \label{fig1}
\end{figure}

\emph{Theories.}---In recent years, several theoretical studies have challenged this conventional belief, introducing the concept of critical topology~\cite{YU20261,Verresen2018PRL,verresen2020topologyedgestatessurvive,Verresen2021PRX,Duque2021PRB,Yu2022PRL,Mondal2023PRB,Yu2024PRB,Prembabu2024PRB,Zhong2025PRB,Huang2025SciPost,Li2025PRB,Rey2025PRB,Zhou2025CP,Cardoso2025PRB,chou2025ptsymmetryenrichednonunitarycriticality,guo2026generalizedlihaldanecorrespondencecritical,deng2026anomalousdynamicalscalingtopological,prembabu2025noninvertibleinterfacessymmetryenrichedcritical,yang2026topologicalquantumcriticalityquasiperiodic,liu2026topologicalsuperradiantuniversalityclasses,TanCP2026,yang2026topologicaltricriticalisinguniversality,chen2026symmetryenrichedcriticalitykennedy,chou2026topologicallyenforcedlifshitzmulticriticality,Keselman2015PRB,Scaffidi2017PRX,JIANG2018753,Parker2018PRB,Thorngren2021PRB,Hidaka2022PRB,Wen2023PRB,Yu2024PRL,Zhang2024PRA,Su2024PRB,Li2025SciPost,Wen2025PRB,Yang2025CP,Flores2025PRL,yang2025deconfinedcriticalityintrinsicallygapless,prembabu2025multicriticalitypurelygaplessspt,Yu2026PRB,xu2026frameworkpredictingentanglementspectra,chou2026topologicallyenforcedlifshitzmulticriticality,chen2026criticaltopologicalphotonicssynthetic,Cheng2026Nature,Lin2026Nature}. To illustrate this phenomenon, consider a one-dimensional quantum spin chain~\cite{Verresen2021PRX,YU20261,Yu2022PRL} described by [Fig.~\ref{fig1}(a)]
\begin{equation}
H=-J\sum_i \sigma_i^x\sigma_{i+1}^x
-g\sum_i \sigma_i^x\sigma_{i+1}^z\sigma_{i+2}^x
-h\sum_i\sigma_i^z,
\end{equation}

where $\sigma_i^\alpha (\alpha=x,y,z)$ denotes a Pauli matrix at site $i$. The three terms describe the nearest-neighbor Ising interaction $J$, the three-spin cluster interaction $g$, and the transverse field $h$, respectively ($J+g+h=1.0$). The model preserves spin-flip $\mathbb{Z}_2$ and time-reversal symmetries and exhibits ferromagnetic, paramagnetic, and SPT cluster phases. For $g=0$, the model reduces to the standard transverse-field Ising chain~\cite{Sachdev_1999}. Tuning $h/J$ drives a continuous Ising transition between a ferromagnetic phase that spontaneously breaks the spin-flip $\mathbb{Z}_2$ symmetry and a symmetry-preserving paramagnetic phase. Conversely, for $h=0$, the model becomes the cluster-Ising chain~\cite{Son_2011}, which undergoes a continuous transition between the ferromagnetic and SPT cluster phases as $g/J$ varies. When both $g$ and $h$ are nonzero, the two QCPs extend into two quantum critical lines [Fig.~\ref{fig1}(b)].
The key question is whether these two critical points belong to the same universality class ? We can address this question from two complementary perspectives. 1) Within the LGW framework, both transitions are associated with spin-flip $\mathbb{Z}_2$ symmetry breaking and are therefore expected to belong to the $(1+1)$-dimensional Ising universality class. 2) Moreover, both models are exactly solvable via the Jordan–Wigner (JW) transformation, yielding critical exponents identical to those of the $(1+1)$-dimensional Ising universality class. By conventional criteria, the two critical points should therefore belong to the same universality class. 

However, the answer is no: the two Ising QCPs possess different topological properties. This distinction becomes apparent in the Majorana representation via the JW transformation [Fig.~\ref{fig1}(a)]. For the critical transverse-field Ising chain ($h/J=1.0$) with open boundary conditions, the boundary Majorana modes are always paired, indicating the absence of topological edge states, consistent with conventional expectations~\cite{landau2013statistical,Sachdev_1999,cardy1996scaling}. In contrast, the critical cluster-Ising chain ($g/J=1.0$) hosts an unpaired Majorana mode at each boundary despite its gapless bulk, resulting in a robust twofold degeneracy. Thus, despite sharing identical critical exponents, these two Ising QCPs have distinct topological properties and cannot be smoothly connected while preserving the protecting symmetries. They must therefore be separated by a transition between critical points—a transition between phase transitions~\cite{Verresen2018PRL,Yu2024PRB,chou2026topologicallyenforcedlifshitzmulticriticality}. We therefore refer to QCPs with nontrivial topological properties—exemplified here by stable topological edge states—as topologically nontrivial QCPs, which provide a paradigmatic example of critical topological phenomena.

Beyond this illustrative example, critical topology has been identified in both strongly correlated quantum critical systems~\cite{Keselman2015PRB,Scaffidi2017PRX,JIANG2018753,Parker2018PRB,Thorngren2021PRB,Hidaka2022PRB,Wen2023PRB,Yu2024PRL,Verresen2021PRX,Duque2021PRB,Yu2022PRL,Mondal2023PRB,Prembabu2024PRB,Zhang2024PRA,Su2024PRB,Li2025SciPost,Wen2025PRB,Yang2025CP,Flores2025PRL,yang2025deconfinedcriticalityintrinsicallygapless,prembabu2025multicriticalitypurelygaplessspt,Yu2026PRB,xu2026frameworkpredictingentanglementspectra,liu2026topologicalsuperradiantuniversalityclasses,deng2026anomalousdynamicalscalingtopological,prembabu2025noninvertibleinterfacessymmetryenrichedcritical} and noninteracting fermionic systems~\cite{Verresen2018PRL,verresen2020topologyedgestatessurvive,Yu2024PRB,Zhang2024PRA,Zhong2025PRB,Zhou2025CP,Cardoso2025PRB,guo2026generalizedlihaldanecorrespondencecritical,chou2026topologicallyenforcedlifshitzmulticriticality,chen2026criticaltopologicalphotonicssynthetic,Cheng2026Nature,Lin2026Nature}, establishing a form of topological phenomena beyond gapped topological phases and topological semimetals, including unconventional topological characterizations~\cite{Verresen2018PRL,Verresen2021PRX,verresen2020topologyedgestatessurvive,Yu2022PRL,yang2026topologicaltricriticalisinguniversality,yang2026topologicalquantumcriticalityquasiperiodic,liu2026topologicalsuperradiantuniversalityclasses} and bulk–boundary correspondences~\cite{Yu2022PRL,Parker2018PRB,Yu2024PRL,Zhong2025PRB,guo2026generalizedlihaldanecorrespondencecritical}, algebraically localized topological edge states~\cite{Verresen2021PRX,Yang2025CP}, anomalous dynamical scaling~\cite{deng2026anomalousdynamicalscalingtopological}, and critical topological states without gapped counterparts~\cite{Thorngren2021PRB,Zhang2024PRA,Li2025SciPost}. For a comprehensive review, see Ref.~\cite{YU20261}
.

\emph{Experiments.}---Despite significant theoretical progress, the experimental observation of critical topology remains challenging. Key issues include accurately locating phase boundaries, characterizing critical behavior, and identifying the nontrivial topology of critical points.

\begin{figure*}[t]
    \centering
   \includegraphics[width=0.85\linewidth]{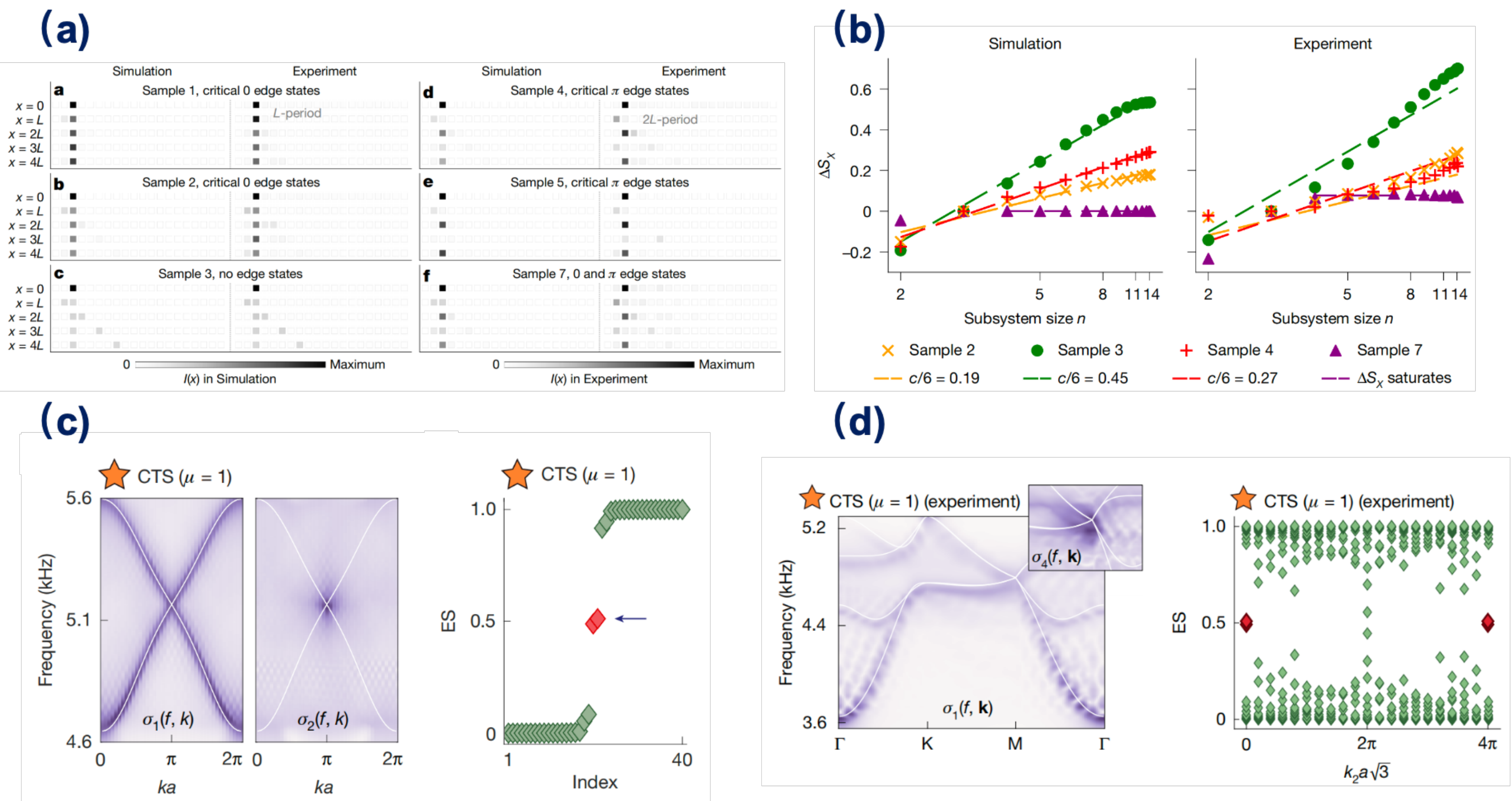}
    \caption{(a) Interference measurements distinguishing critical edge modes at quasienergies $0$ and $\pi$. (b) Entanglement-entropy measurements. The entanglement coefficient extracted from sample 3 equals the sum of those obtained from samples 2 and 4, providing direct experimental evidence for topology-enforced multicriticality. (c) Experimental characterization of critical topological states (CTSs). The left and right panels show the measured singular values $\sigma_1(f,k)$ and $\sigma_2(f,k)$, respectively. The former yields the phonon band dispersion, while the latter characterizes band degeneracies. The measured entanglement spectrum of the one-dimensional CTS is also shown. (d) Experimentally measured band dispersion of the CTS, with the measured $\sigma_2(f,k)$ shown in the inset. The right panel shows the measured entanglement spectrum of the two-dimensional CTS.Panels (a) and (b) are adapted from Ref.~\cite{Cheng2026Nature}, and panels (c) and (d) from Ref.~\cite{Lin2026Nature}.}
    \label{fig2}
\end{figure*}

Recently, critical topology has been experimentally observed by two independent research groups using classical-wave platforms. The first group~\cite{Cheng2026Nature} investigated a one-dimensional Floquet system using acoustic metamaterials. Specifically, they employed the periodically driven, chiral-symmetric Floquet Su–Schrieffer–Heeger (SSH)-$\alpha$ chain~\cite{Zhou2025CP,Cardoso2025PRB} and mapped its dynamics onto sound propagation through an engineered waveguide array [Fig.~\ref{fig2}(a) and (b)]. The experimental sample comprised $20$ waveguides, with sound waves excited at approximately $9.8$ kHz. By varying the model parameters, the researchers fabricated samples representing different points in the phase diagram. Sound sources were placed inside the array to probe bulk modes and near its physical boundaries to detect localized edge responses. Measurements of the sound-field distributions after multiple Floquet periods, combined with tight-binding simulations, enabled the reconstruction of the quasienergy spectrum and the identification of spectral features along distinct phase boundaries and at their intersections. Because sound-intensity measurements alone cannot distinguish edge modes at quasienergies $0$ and $\pi$, the researchers performed additional interference measurements. By comparing interference patterns across different samples, they distinguished the two types of critical edge modes associated with topologically distinct critical lines. They also investigated the scaling of entanglement entropy with subsystem size. At the intersection of two topologically distinct critical lines, the logarithmic-scaling coefficient obeyed an additive relation, equaling the sum of the contributions from the two lines. Together, these complementary results provide experimental evidence for critical topology and topology-enforced multicriticality~\cite{Verresen2018PRL,Yu2024PRB,chou2026topologicallyenforcedlifshitzmulticriticality}.

Meanwhile, another group~\cite{Lin2026Nature} experimentally characterized critical topological states using a newly developed entanglement-spectrum measurement technique for phononic systems~\cite{Lin2024NC} and extended the investigation to two dimensions [Fig.~\ref{fig2}(c) and (d)]. Because critical topological states lack a bulk energy gap, conventional methods developed for gapped systems are generally insufficient to characterize their topology. The entanglement spectrum, by contrast, provides a valuable diagnostic of topology in critical systems~\cite{Yu2024PRL,Zhong2025PRB,guo2026generalizedlihaldanecorrespondencecritical}. Motivated by this observation, the researchers experimentally characterized critical topological states in both one and two dimensions and demonstrated topology-enforced multicriticality. These results demonstrate the possibility of topologically localized boundary states embedded in the bulk continuum, with potential applications in acoustic and optical systems.

\emph{Outlook.}---Critical topology has emerged as a form of quantum criticality beyond the Landau paradigm, challenging conventional approaches to classifying universality classes by introducing topological properties as an additional criterion. Despite substantial progress over the past five years, several fundamental questions remain open. Theoretically, can we establish a general criterion for the emergence of critical topology? Can a unified theoretical framework be extended beyond one-dimensional quantum many-body systems? What efficient numerical methods can be developed to investigate critical topological states in higher-dimensional strongly correlated systems? Experimentally, research has so far focused primarily on single-particle critical topology. An important next step is to realize and characterize many-body critical topology in quantum systems using platforms such as Rydberg atom arrays, superconducting quantum processors, and quantum materials.

Major advances in statistical physics often arise from re-examining fundamental concepts. The theory of critical phenomena developed by Landau, Ginzburg, and Wilson, together with the topological phase transitions pioneered by Kosterlitz and Thouless, has now inspired new discoveries in artificial acoustic metamaterials. These experimental observations mark a beginning rather than an endpoint for research on critical topology, and further advances in this direction are anticipated.

\textit{Acknowledgement}:
We thank Hai-Qing Lin, Limei Xu, Jiangbin Gong, Baile Zhang, Jian-Hua Jiang, Sheng Yang, Yuxuan Guo, Shao-Kai Jian, Yi-Zhuang You, Da-Chuan Lu, Fuxiang Li, Long Zhang, Po-Yao Chang, Kuang-Hung Chou, Haohua Wang, Chao Song, Longwen Zhou, Zheyu Cheng, Zhi-Kang Lin, Franco Nori, Saranesh Prembabu, Yunqing Zheng, Rui Wen, and Linhao Li for fruitful collaborations and helpful discussions.
X.-J. Yu was supported by the National Natural Science Foundation of China (Grant No.12405034) and a start-up grant from Eastern Institute of Technology, Ningbo. 

\bibliographystyle{apsrev4-2}
\let\oldaddcontentsline\addcontentsline
\renewcommand{\addcontentsline}[3]{}
\bibliography{main.bib}

\end{document}